\documentclass[11pt]{article}

\usepackage[a4paper,margin=1in]{geometry}
\usepackage[T1]{fontenc}
\usepackage[utf8]{inputenc}
\usepackage{lmodern}
\usepackage{microtype}

\usepackage{amsmath,amssymb}
\usepackage{booktabs}
\usepackage{graphicx}
\usepackage[font=small,labelfont=bf]{caption}
\usepackage{xcolor}
\usepackage{tikz}
\usetikzlibrary{arrows.meta}
\usepackage{placeins}
\usepackage[hyphens]{url}

\usepackage{authblk}

\usepackage{natbib}
\usepackage[colorlinks=true,linkcolor=blue!60!black,citecolor=blue!60!black,urlcolor=blue!60!black]{hyperref}

\newcommand{\answerYes}[1]{\textcolor{blue}{#1}} 
 
\newcommand{\answerNA}[1]{\textcolor{gray}{#1}}

\title{Perceived Trust and Transparency in AI-Altered vs. Human-Written LinkedIn Posts}

\author[1]{Thomas Christian Joos}
\author[1]{Azza Bouleimen}
\author[1]{Anik\'{o} Hann\'{a}k}
\author[1]{Nicol\`{o} Pagan}
\affil[1]{University of Zurich, Zurich, Switzerland\\
\texttt{thomaschristian.joos@uzh.ch, nicolo.pagan@uzh.ch, azza.bouleimen@uzh.ch, hannak@ifi.uzh.ch}}

\date{\today}

\begin{document}

\maketitle

\begin{abstract}
People increasingly draft a social media post themselves and then let an AI model refine it (AI-altered), a common but understudied middle case between fully human-written and fully AI-generated content. We ask whether this changes how trustworthy and transparent readers perceive the post to be, and whether they can even tell. We first analyze real ChatGPT conversations (WildChat) to find how people actually alter social media posts, which we use to design realistic alteration prompts. We then run a pre-registered, within-subjects online experiment (n=214) in which participants rate unlabeled human-written and AI-altered LinkedIn-style posts for perceived trustworthiness and transparency, and separately judge whether each post was AI-altered. We find that AI-alteration itself has no detectable effect on either perceived trustworthiness or transparency (both estimates near zero, not significant). At the same time, readers did not reliably detect AI-alteration (49\% accuracy, near chance) and defaulted to assuming human authorship when unsure. 
However, our analysis shows significant differences in transparency and trustworthiness ratings across the specific post, the AI model, as well as the prompt used to alter it. For instance, more detailed prompts raise both perceived trustworthiness and transparency but also make the alteration more detectable. Surprisingly, the prompt that shifted ratings most also asked the model to add one new idea absent from the original post: readers nonetheless rated this AI-invented, unverifiable content as more trustworthy and transparent. 
\end{abstract}

%
%
\section{Introduction}
People often say they would trust a message less once they suspect AI wrote it: in an experiment on real, entirely human-written Airbnb profiles, hosts explicitly labeled as using an AI writing assistant were rated less trustworthy than hosts believed to write their own \citep{jakesch2019}.
Yet this suspicion may not track anything real: across six later experiments with over 4{,}600 participants, people could not reliably tell genuinely AI-generated self-presentations from human-written ones in professional, hospitality, and dating contexts \citep{jakesch2023}, so when distrust of ``AI-sounding'' text does occur, it likely falls on the wrong content as often as the right one.
Nor is disclosure's effect itself settled: a larger, more recent experiment that labeled real AI-generated misinformation and real human-written content alike as AI-generated found no significant effect of the label on perceived accuracy or credibility \citep{li2024}, where \citet{jakesch2019}'s comparable labeled condition had found a significant drop.
Whether, and under what conditions, knowing or suspecting AI involvement changes how content is judged is thus still unresolved, and this paradox plays out wherever people write for others online.
And such AI-assisted writing is becoming the default: about half of marketers already use AI to create or optimize content \citep{surveymonkey2025}, and by the end of 2024, almost 40\% of the US working-age population reported using generative AI, and 78\% of organizations reported using AI that year (up from 55\% in 2023), an adoption curve faster than the computer or the Internet's \citep{bick2024,maslej2025}.  
That capability cuts both ways: the same models that can polish an existing post can just as easily generate a persuasive one from scratch, at a scale no individual writer could match \citep{salvi2025,chergarova2025}.


Most of this writing assistance is not full generation but alteration: a person drafts a post, perhaps a rough note about a product launch or a work update, and lets a model brainstorm, refine the language, adjust the tone, or extend it, changing how something is said but not what is said (\textbf{AI-altered}), rather than generating a post entirely from a topic or brief instruction with no human draft to begin with (\textbf{AI-generated}; \citealp[p.~155]{mindner2023}).
Yet the paradox above, whether and when knowing or suspecting AI involvement changes how content is judged, has so far been tested mainly by manipulating AI-authorship labels or beliefs~\citep{jakesch2019, li2024}. In this work, we test whether \textbf{AI-alteration} changes how trustworthy and transparent readers perceive a post to be, and whether they can tell the difference in the first place. 
This leads to the following research question.

\begin{itemize}
    \item \textbf{RQ:} Are AI-altered social media posts perceived as trustworthy and transparent as their corresponding human-written ones? If yes, under which prompting conditions?
\end{itemize}

A pre-registered experiment addresses the RQ directly, combining three steps.
First, we analyze the WildChat dataset \citep{zhao2024} to see how people actually use ChatGPT to alter social media posts, informing the prompting styles used in the next step.
This analysis also shows LinkedIn is the most common target platform for such requests, and professional networks are a high-stakes case in their own right: posts there are tied to a real name, a real employer, and colleagues who can check what is claimed \citep{guillory2012}, and the platform is projected to grow from 644 million users in 2019 to 942 million by 2028 \citep{statista2023}.
We then gather real LinkedIn posts, alter them with two LLMs and four prompts, and compare originals to alterations with text metrics.
Building on this, a pilot and the pre-registered main experiment let participants rate perceived trustworthiness and transparency of AI-altered LinkedIn posts, and later separately judge, for each post, how likely it is human-written, testing whether readers can detect AI-alteration at all, which can affect how much they trust it (see \emph{Method} below).

Our hypotheses are the following:
\begin{itemize}
    \item \textbf{H1 (Trustworthiness):} AI-altered social media posts are perceived as \textit{more} trustworthy than human-written social media posts.
    \item \textbf{H2 (Transparency):} AI-altered social media posts are perceived as \textit{more} transparent than human-written social media posts.
\end{itemize}
We expect AI-altered posts to be rated higher on both scales, as they are more clearly written than human-written ones.
We also expect this effect to outweigh people's skepticism towards AI-altered posts, which may come up if they detect that a text was created with the help of AI.
We expect however that, for LinkedIn-style posts, the AI-detection task is not trivial, since the platform is used by professionals who tend to write clearer posts, which could be more easily confounded with those altered with an AI tool.
Additionally, we explore which LLM model and prompt style have a greater impact on perceived trustworthiness and transparency.

Two results anchor the paper.
First, altering a post with AI does not, by itself, move perceived trustworthiness or transparency in any significant way: these perceptions can still be raised or lowered by which model altered a post and which prompt was used, even though AI-alteration itself is invisible on average. 
Second, this null result has a straightforward cause: the experiment revealed that participants could barely tell an altered post from an unaltered one, and when in doubt, assumed a human had written it. Therefore, an alteration that raises no suspicion just gets judged as trustworthy and transparent as when a person had written the whole post.
A further, more practical finding is that richer prompts raise both perceived trustworthiness and transparency at the cost of making the alteration itself easier to spot, including when the prompt explicitly asks the model to sound less like AI. The richest prompt also asked the model to add a new idea the post did not originally contain (content invented by the AI rather than supplied or verified by the human author) and readers still rated it as more trustworthy and transparent.
We use these results to situate AI-alteration on a four-stage spectrum of AI involvement in social media writing, distinct from fully AI-generated content.

\section{Related Work}

We cover AI-alteration and our two core constructs, perceived trustworthiness and transparency. Further background (generative AI and writing, AI-mediated communication, AI adoption on social media, and platform-level trust studies) is in the Appendix.

\subsection{AI-Alteration}

As introduced above, a person's draft can be refined by AI rather than generated by it from scratch. The literature describes this case with different terms, including AI-modified \citep{arnold2025}, AI-rephrased \citep{mindner2023}, and AI-assisted \citep{xie2025}, which we unify as \textbf{AI-alteration}: an AI rewriting existing human input \citep{mindner2023,xie2025} without changing its content \citep{arnold2025,barrot2025} or making changes too significant \citep[p.~302]{furey2025}, so it makes only linguistic or stylistic enhancements.
AI-alteration and AI-generation form a spectrum rather than a binary (\citet[p.~2]{xie2025} call it a ``grey area between these two extremes''), separated in prior work by creation process (human-written/edited, AI-edited/written; \citealp{tian2025}), amount of change \citep[p.~7]{barrot2025}, presence of a source text \citep{mindner2023}, or degree of human contribution \citep{xie2025}.
We correspondingly treat a text as AI-generated when human input is minimal (e.g., only a topic), from which the AI generates a large text \citep{mindner2023,barrot2025,xie2025}.

A related concept, AI-Mediated Communication (AI-MC), is communication ``in which a computational agent operates on behalf of a communicator by modifying, augmenting, or generating messages'' \citep[p.~90]{hancock2020}, studied for AI-authored Airbnb profiles and trust \citep{jakesch2019}, AI-generated replies reducing social attraction \citep{mieczkowski2021}, and AI-mediated video reducing trust, most strongly alongside unaltered video \citep{navajas2026}.
Our setting differs in that communication runs to an entire network, not one recipient.
At the far end of the spectrum, social networks populated entirely by AI agents are emerging as a research object in their own right \citep{jiang2026,demarzo2026,lin2026}; these studies describe the collective behavior and dynamics of such agent communities rather than human readers' perceived trust or transparency, which is the gap this paper addresses instead.

\subsection{Trustworthiness and Transparency}
Trustworthiness and transparency, the two constructs this paper measures, require working definitions before they can be operationalized.
Trust is ``the willingness of a party to be vulnerable to the actions of another party'' \citep[p.~712]{mayer1995}, with Ability, Benevolence, and Integrity (ABI) as antecedents \citep[p.~719]{mayer1995}; adapted to the perceiving reader, our working definition is: \textbf{a post is trustworthy if you can trust it because the source seems honest and the content credible.}
Transparency is usually defined for organizations as making enough information available to understand what was done \citep{rawlins2008,florini2007}. Adapted to a post: \textbf{it is transparent if it openly shares enough information to understand and judge what was decided, done, or how the post was created.}
We measure \emph{perceived} trustworthiness/transparency, since a reader scrolling a feed does not verify a post directly \citep{jakesch2019}, using a reduced ABI item scale \citep{ma2017,jakesch2019} plus a confirmation item.
A key prior finding links the two concepts directly: the ``Replicant Effect'', where suspicion of AI authorship reduces trust even when all content is human-written \citep{jakesch2019}, though AI labels alone do not always reduce credibility \citep{li2024}.
Yet no validated scale exists for perceived transparency of a post, and platforms like LinkedIn, where AI-alteration is already common, remain unstudied in this respect; we build a transparency scale from these definitions and test both scales on real, unlabeled, genuinely AI-altered LinkedIn posts.

\section{Method}

We test the RQ with a controlled online experiment in which participants rate human-written and AI-altered posts for perceived trustworthiness and transparency, and separately later judge which is which.
This section proceeds in two parts: first a formative analysis that fixes the AI-alteration conditions, two AI models and four prompts, from real-world evidence, then the main experiment itself, this paper's central study.

\subsection{Formative Analysis: Deriving the AI-Alteration Conditions}
We proceed in two steps: mining real ChatGPT conversations for how people already alter social media posts, then screening candidate models and prompts on a corpus of real LinkedIn posts.

\subsubsection{Real-World Alteration Patterns}
WildChat is a dataset of almost one million real, ``in the wild'' ChatGPT conversations from 2023--2024 with more than 2.5 million turns, collected by offering free access in exchange for logging consent \citep[p.~3]{zhao2024}. Around 53\% of turns are in English, and ``assisting/creative writing'' is the most common first-message request type ($\sim$62\%; \citealp[p.~3]{zhao2024}).

To find real-world AI-altering prompts for social media posts within this dataset, we build a small pipeline. English turns above a minimum length are first pre-filtered with keywords for social media (e.g., LinkedIn post, tweet) and for altering (e.g., rewrite, rephrase), while generation requests and unrelated text types are excluded; each remaining conversation is then classified with a local LLM (\texttt{Qwen2.5:14b} via Ollama), prompted to confirm that the user is asking to alter, not generate, an existing post. A conversation is kept only when the classifier and the rule-based filter agree.

Of 837{,}989 collected conversations, 544{,}445 had at least one qualifying English turn, but AI-alteration itself turned out to be rare: only 53 conversations requested altering an existing social media post. Annotating these by area, platform, and prompting style shows a fairly consistent picture: the most common areas were marketing ($n=14$) and news ($n=11$), the most common platform was LinkedIn ($n=17$, ahead of Instagram, Reddit, X/Twitter, and Facebook), and the most common prompt types asked to rewrite as a social media post, rewrite it in a specific way (e.g., tune the tone or style, such as making it funnier or more formal), or take a persona (e.g., write as a specific role, such as a marketer) ($n=8$--$10$ each) (See Table~\ref{tab:promptgroup} in the Appendix). This small analysis already fixes two of our three choices: we focus on LinkedIn, the most common target platform, and combine these patterns, generically fitting a post to a given platform and tuning it along some characteristic, into the four alteration prompts used in the main experiment (see \emph{Stimuli}, below).

\subsubsection{Model Screening}
For the third choice, which AI models to use, we first collect our own corpus of 99 real LinkedIn posts via Google search restricted to before 2020 (e.g., \texttt{site:linkedin.com/posts news before:2020-01-01}), predating consumer LLMs (GPT-3.5 launched 2022) and therefore not yet AI-influenced. These posts are then altered with four candidate models (\texttt{Qwen2.5:7b-instruct}, \texttt{Gemma3:4b}, \texttt{GPT-5.2}, \texttt{Claude Sonnet 4.6}) and three WildChat-derived prompting styles, giving a 1{,}188-post screening corpus.

For every original/altered pair, we compute the \% change across five groups of text metrics: \textit{lexical diversity} (unique words, total words, Type-Token Ratio, letter count, average word length), \textit{sentence structure} (count, words/characters per sentence), \textit{formatting} (special characters, punctuation, line breaks, paragraphs), \textit{social-media features} (emoji/hashtag count and density, uppercase ratio), and \textit{sentiment} (score and change), plus cosine similarity between the original/altered pairs (using \texttt{all-MiniLM-L6-v2} embeddings).

Across the screening corpus, AI-alteration mostly expands and reformats posts: words, letters, sentences, special characters, punctuation, and paragraphs all increase, while content-level metrics stay comparatively stable (TTR, uppercase ratio, sentiment, mean cosine similarity). Averaged across the eight metrics in Table~\ref{tab:screeningmodels} (Appendix), \texttt{Qwen2.5:7b-instruct} changed posts the least (mean absolute change $+29.8$\%) and \texttt{Claude Sonnet 4.6} the most ($+76.4$\%), while \texttt{Gemma3:4b} shifted sentiment score the most ($+17.0$\%). \texttt{Claude Sonnet 4.6} and \texttt{Gemma3:4b} are thus the two most differentiated models; together with the four prompts derived above, they become the AI-alteration conditions carried into the main experiment described next.

\subsection{Experimental Design}
Using the two models and four prompts fixed above, we ran a controlled online experiment hosted on an institutional Qualtrics instance, styled to resemble LinkedIn, in which each participant sees posts one by one, as in a real feed, and rates each one of them on multiple items.
After the information and consent page, the experiment has three consecutive parts: (i) rating posts for trustworthiness and transparency, (ii) classifying the posts of part (i) into human-written or AI-altered, and (iii) a short follow-up questionnaire, followed by debriefing (see \emph{Participant Instructions} in the Appendix for screenshots).

\subsubsection{Stimuli}
The posts used in the experiment consist in six human-written posts and their AI-altered versions: six real LinkedIn posts from before 2020, a separate, non-overlapping set from the screening corpus in \emph{Model Screening}, above, collected via the same Google-search method and covering the major WildChat topic areas identified above (four from marketing/news, one from business/technology, one from a personal topic).
Each of the six was altered with the two models selected above (\texttt{Claude Sonnet 4.6}, \texttt{Gemma3:4b}) using the four prompts derived there: 
\begin{enumerate}
    \item Refine this post.
    \item Refine this post to fit LinkedIn, but make it not sound like AI.
    \item Refine this post to fit LinkedIn.
    \item Refine this post to fit LinkedIn. Also insert one additional idea that was not contained yet (wasn't already) in the post, making it sound more engaging and exciting.
\end{enumerate}

This gives 48 AI-altered posts ($6 \times 2 \times 4$) plus the six originals, 54 stimuli in total.
\texttt{Claude Sonnet 4.6} was run in Claude Desktop, incognito, on a fresh account (no memory of prior posts), resembling how people use it; \texttt{Gemma3:4b} ran locally via Ollama.
Where a model offered several versions, only the first is used; proposed links or images are removed, author names are redacted to ``Social Media User'' to avoid bias, and posts are displayed with custom CSS in Qualtrics to resemble a LinkedIn feed.
An information page introduces the task and asks for participant's consent without mentioning AI usage, and posts are never labeled as AI-altered or human-written: disclosing this would make participants think about authorship and bias their ratings, consistent with prior findings \citep{jakesch2019}.

\subsubsection{Perceived Trustworthiness and Transparency}
To measure participants' perception of trustworthiness and transparency, we compile a set of items to rate, following prior research \citep{mayer1995}.

\noindent \emph{Trustworthiness} is rated with the following items:

\begin{enumerate}
\item The author of this post knows about the topic. (Ability)
\item The author thought about the impact on readers when writing this post. (Benevolence)
\item The author seems honest and genuine. (Integrity)
\item I perceive this post as trustworthy. (Confirmation item)
\end{enumerate}
The first three items adapt the ABI scale to the social media setting; the confirmation item asks participants to rate trustworthiness directly, letting us check whether the ABI-based scale aligns with a holistic judgment.

\noindent \emph{Transparency} is rated with the following items:
\begin{enumerate}
\item It is clear what the information in this post is based on (e.g., the author's own experience, specific evidence, or common knowledge).
\item I perceive this post as transparent. (Confirmation item)
\item I find this post clear and understandable. (Clarity item)
\end{enumerate}
The first item operationalizes our transparency definition (see \emph{Trustworthiness and Transparency} in Related Work) as something a reader can judge directly from the post.
The second item is a confirmation item analogous to the trustworthiness items.
The third item is exploratory: we test with Cronbach's $\alpha$ whether clarity belongs to the same construct as transparency (see Results).

\subsubsection{AI-Detection}
In part (ii), participants see the same posts again and rate: ``I think this post was written entirely by a human, without the use of AI (language model tools)''.
This detection task comes in its own, later part, separate from the trustworthiness/transparency ratings, since thinking about authorship while rating could otherwise bias participants against posts they suspect are not human-written.

\subsubsection{Post-Survey / Follow-up Questions}
At the end of the experiment, participants report what they relied on when rating posts (formatting, content, writing style, etc.) and answer a few exploratory questions on LinkedIn use and perception, to help interpret the main ratings.

\subsubsection{Scale}
All scales use a five-point Likert scale (1 = Strongly disagree to 5 = Strongly agree).
In parts (i) and (iii) of the experiment, we include, in addition to the neutral midpoint, a separate ``don't know'' option, since being neutral and being unable to judge are different responses \citep{wetzelhutter2020} and are treated separately in the analysis. In part (ii) (AI detection) they collapse into one ``Don't know / can't say'' midpoint, since, in this task, being undecided and not knowing are the same.

\subsubsection{Procedure}
Every participant rates four posts in a random order within each part: 1 human-written post and 3 AI-altered ones. The human-written post is randomly sampled from the six human-written originals. The AI-altered posts are drawn from one of the eight model$\times$prompt combinations and kept consistent across parts 1 and 2.
This way, each participant never gets to rate both the human and the AI-altered versions of the same post. 
The items within each rating task are also randomized.
Participants are not told that some posts are AI-altered until debriefing.
Attention checks appear as extra items on two posts per part (e.g., ``Please select `1 - Strongly disagree'.''). Failures are recorded and used for exclusion.

\subsubsection{Participants and Recruitment}
We recruited participants from Prolific who were fluent in English and used LinkedIn at least monthly. We compensated them with 7 £/hour. The experiment required around 13 minutes to be completed.
We excluded participants who failed the attention checks, did not finish, or had other data issues. 

\subsubsection{Analysis Plan and Statistical Model}
Prior to the final data collection, we pre-registered the statistical models, reliability checks, exclusion rules, hypotheses, and planned sample size on OSF to guard against Researcher Degrees of Freedom \citep{simmons2011}.

As described in \emph{Procedure}, each participant rates multiple posts, and each post is rated by multiple participants, so ratings are non-independent along two dimensions simultaneously. We therefore adopt, for this study, a Linear Mixed-Effects Model (LMEM) including crossed random intercepts for participant and post. Specifically, our statistical model is the following:
\begin{equation}
    \mathit{rated\_attribute} \sim \mathit{condition} + (1|\mathit{participant}) + (1|\mathit{post})
    \label{LMEM}
\end{equation}
where the $rated\_attribute$ is transparency or trustworthiness, and $condition$ represents where the post is human-written or AI-altered, as the fixed effect for H1 and H2.

Before averaging trustworthiness items (three ABI indicators plus a general item) into one score, we checked internal consistency with Cronbach's $\alpha$ ($\geq 0.70$ = reliable). We checked transparency the same way, comparing $\alpha$ with and without the clarity item to decide whether it belongs to that scale. The trustworthiness score averages the four trustworthiness items and the transparency score averages the three transparency items.


Before the main run, we ran a pilot study using the same procedure and materials with $N=20$ participants. Descriptively, the pilot pointed toward AI-altered posts scoring higher than human-written ones on both trustworthiness and transparency, a direction the main study did not confirm (\emph{No Difference of AI in Ratings}). We estimated the required sample size with a power simulation (\texttt{simr} package in R \citealp{green2016}) matching the LMEM above, targeting power of $0.80$ at a significance level of $\alpha=0.05$ for an effect size of $d=0.20$ estimated from the pilot. The simulation indicated a minimum of 190 participants for detecting an effect in trustworthiness and 220 in transparency.

\section{Results}

\subsection{Sample and Data Validity.}
200 new participants completed the study in under an hour (median 11min 33s, average reward 7.30 GBP/hour), yielding 9{,}117 data points (7{,}760 in part 1, 1{,}357 in part 2, 2{,}568 follow-up answers, including examples and attention checks) across 201 completed responses, one participant having submitted twice.
We excluded this duplicate, one participant who reported misreading the part-2 scale, and five who failed attention checks, leaving 194 valid main-study responses (214 with the pilot).
We did not exclude by completion time (no principled cutoff) and found no participant giving the same answer throughout (per-participant SD range 0.496--1.800, mean 1.089).

Cronbach's $\alpha$ is good for trustworthiness (four items, $\alpha=0.851$) and for transparency with the clarity item (three items, $\alpha=0.804$), but only acceptable without it (two items, $\alpha=0.736$), so the clarity item belongs in the transparency scale.\footnote{$\alpha>0.90$ excellent, 0.80--0.90 good, 0.70--0.80 acceptable, $<0.70$ poor.}

\subsection{Manipulation Check: Did the Alteration Work?}
Before turning to the ratings, we check the experiment stimuli themselves with the same eight text metrics as in \emph{Model Screening} (word count, special characters, punctuation, paragraphs, line breaks, TTR, uppercase ratio, sentiment score), plus cosine similarity to the original.
The same metrics, now on our 48 main-experiment AI-altered posts (6 posts $\times$ 2 models $\times$ 4 prompts; full per-metric breakdown by model and by prompt in Table~\ref{tab:experimentmetrics}, and by post in Table~\ref{tab:perpost}, Appendix), show the shortest post (Post 5, two sentences) changed most (mean absolute change $+67.62$\%, consistent with short posts being expanded) and Post 1 changed least ($+14.19$\%).
By model, Claude Sonnet 4.6 changes posts more (mean absolute change $+34.24$\%) and more similarly to the original (cosine 0.884) than Gemma3:4b ($+19.18$\%, cosine 0.837), which instead shifts sentiment far more ($+29$\% vs.\ $+2$\%).
By prompt, the basic ``Refine this post.'' is gentlest on every metric ($+10.46$\%, cosine 0.884) and the additional-idea prompt strongest ($+56.30$\%, cosine 0.826), as expected since it explicitly requests new content.
We relate these metrics to the ratings in the Appendix (\emph{Post-Level Metrics vs.\ Ratings Correlations}).

To turn these averages into a statistical claim, we test, across the 48 stimuli, whether each metric's percentage change differs from zero (one-sample $t$-test, with Wilcoxon signed-rank as a robustness check given the skew of percentage-change data; Table~\ref{tab:metricstats}).
Word count, special characters, punctuation, paragraphs, and line breaks all increase reliably, with small-to-medium effect sizes ($d = 0.39$--$0.59$).
TTR does not change reliably ($p = 0.617$), confirming that lexical diversity stays stable.
Sentiment score increases reliably ($d = 0.42$) but with a large spread (SD $= 37$), so the direction is consistent but the magnitude varies a lot by post.
Uppercase ratio is borderline: significant under the $t$-test ($p = 0.044$) but not under the more conservative Wilcoxon test ($p = 0.073$), so we treat it as inconclusive.
Cosine similarity to the original (not a change score, so not tested against zero) has mean $0.861$ (median $0.869$, SD $0.065$) across the 48 stimuli, consistent with the model-screening corpus.
In short, the alteration worked as intended: it reliably changed form (length, formatting) while leaving content-level signals (TTR, similarity) largely intact, giving us genuinely, and validly, AI-altered stimuli to rate next.

\begin{figure}[!htbp]
\centering
\includegraphics[width=.9\textwidth]{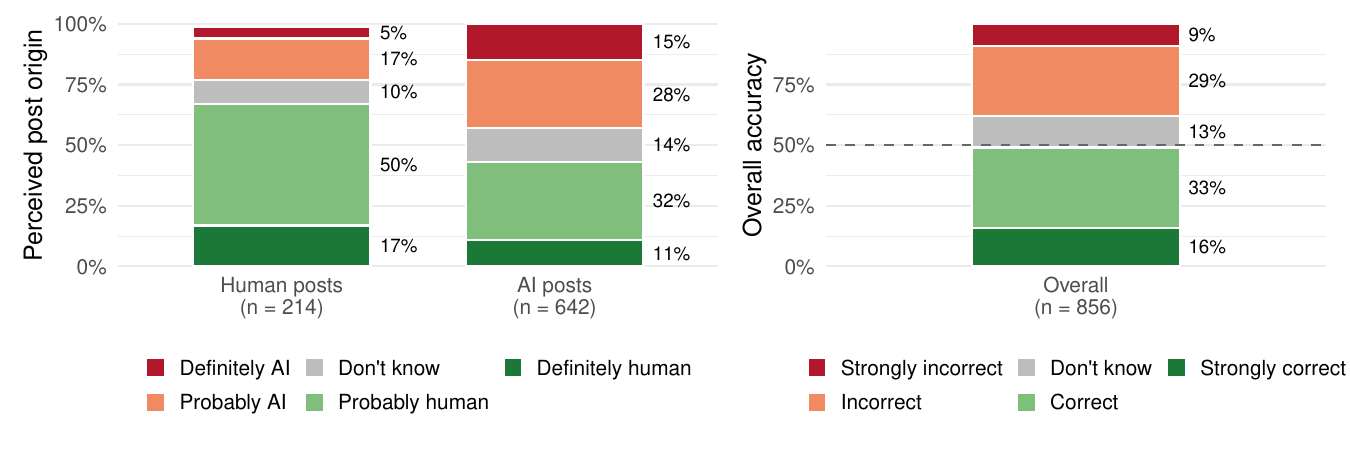}
\caption{Can people tell AI-altered posts from human ones? Distribution of Detection Judgments by Post Type (left) and Overall Accuracy Graded by Confidence (right).}
\label{fig:detection}
\end{figure}

\begin{table}[!htbp]
\centering
\small
\begin{tabular}{lrrr}
\toprule
Metric ($n=48$) & Mean \% & $t$ & Cohen's $d$ \\
\midrule
Word count & $+41.7$ & 2.69* & 0.39 \\
Special characters & $+97.3$ & 3.77*** & 0.54 \\
Punctuation & $+66.8$ & 2.92** & 0.42 \\
Paragraphs & $+91.8$ & 4.12*** & 0.59 \\
Line breaks & $+64.0$ & 3.69*** & 0.53 \\
TTR & $+0.7$ & 0.50 & 0.07 \\
Uppercase ratio & $+9.0$ & 2.07* & 0.30 \\
Sentiment score & $+15.8$ & 2.94** & 0.42 \\
\bottomrule
\end{tabular}
\caption{One-sample tests of each metric's \% change against zero across the 48 main-experiment stimuli. $*p<.05$, $**p<.01$, $***p<.001$ ($t$-test); uppercase ratio is not significant under a Wilcoxon signed-rank test ($p=.073$).}
\label{tab:metricstats}
\end{table}

\subsection{No Effect on Perceived Trustworthiness and Transparency.}

Table~\ref{tab:h1h2} shows the LMEM results on the perceived trustworthiness and transparency: for both outcomes, the AI-vs-human difference is close to zero, not significant, with narrow confidence intervals (CIs); participants rated the transparency and trustworthiness of AI-altered posts almost the same as human-written ones. We find no evidence to support H1 and H2.

\begin{table}[!htbp]
\centering
\small
\begin{tabular}{lrrr}
\toprule
Outcome & Estimate & 95\% CI & $p$ \\
\midrule
Trustworthiness & $+0.016$ & $[-0.093, 0.125]$ & .777 \\
Transparency & $+0.013$ & $[-0.113, 0.139]$ & .840 \\
\bottomrule
\end{tabular}
\caption{LMEM estimates for AI-altered vs.\ human-written posts (human-written = intercept).}
\label{tab:h1h2}
\end{table}

The following subsections unpack this null: whether readers could actually tell AI-altered posts apart, and whether specific models or prompts moved ratings even though the average difference did not. How much of the variation in ratings traces to the individual post is reported in the Appendix (\emph{Exploratory Analyses: Post-Level Variance}), since with only six posts spanning two topic areas we cannot draw generalizable conclusions from it.


\subsection{Readers Do Not Notice the Alteration.}

Figure~\ref{fig:detection} shows how accurately participants detected which posts are AI-altered and which are human-written.
Overall, participants were correct in 49.1\% of all 856 judgments, incorrect in 37.7\%, and chose ``don't know'' in 13.2\% (56.5\% vs.\ 43.5\% if ``don't know'' is excluded), only slightly better than guessing.
Accuracy is asymmetric: human-written posts were correctly identified 67.3\% of the time, but AI-altered posts only 43.0\%.
AI-altered posts were labeled human about as often as AI (275 vs.\ 276 times), while human-written posts were labeled human three times as often as AI (144 vs.\ 48), showing participants assume human authorship by default. This is confirmed by confidence levels in Figure~\ref{fig:detection}, where confidently wrong ``definitely'' judgments were twice as common on AI-altered posts (10.6\% ``definitely human'') as on human-written posts (5.1\% ``definitely AI'').

Detection accuracy on AI-altered posts also varies by model and prompt style as shown in Table~\ref{tab:detbycell}.
Averaged over prompts, posts altered with \texttt{Claude Sonnet 4.6} were recognized as likely to be AI more often (48.9\%) than posts altered by \texttt{Gemma3:4b} (37.2\%); the models which altered the posts that got the lower trustworthiness ratings is at the same time more difficult to detect.
When averaged over models, the basic prompt to refine a post was hardest to judge correctly (38.7\%), and the prompt that adds an additional idea was easiest (48.6\%); the prompt asking the model not to sound like AI did not make posts less detectable (44.6\%).

Table~\ref{tab:detbycell} breaks this down into the full model$\times$prompt matrix.
The pattern from the marginals holds within each model: \texttt{Gemma3:4b} is harder to detect than \texttt{Claude Sonnet 4.6} at every one of the four prompts.
The ``not sound like AI'' prompt fails at its stated goal for both models (it is not the least-detected prompt for either), and for \texttt{Claude Sonnet 4.6} it is tied for the \emph{most}-detected (52.6\%, decided-only 58.0\%).

\begin{table}[!htbp]
\centering
\small
\begin{tabular}{lrrr}
\toprule
Prompt & Claude & Gemma & Both \\
\midrule
Refine & 44.8 (50.6) & 32.1 (40.6) & 38.7 (46.1) \\
Not sound like AI & 52.6 (58.0) & 37.8 (45.3) & 44.6 (51.4) \\
Fit LinkedIn & 46.4 (50.6) & 35.7 (45.5) & 41.1 (48.3) \\
Additional idea & 52.8 (60.3) & 44.1 (50.0) & 48.6 (55.3) \\
\midrule
All prompts & 48.9 (54.5) & 37.2 (45.3) & 43.0 (50.0) \\
\bottomrule
\end{tabular}
\caption{Share of AI-altered posts detected as AI (\%), by prompt and model, all judgments (excluding ``don't know'' in parentheses). Both = both models combined.}
\label{tab:detbycell}
\end{table}


\subsection{The Model and the Prompt Still Matter.}

\subsubsection{AI Models and Prompts}
\begin{figure}[!htbp]
\centering
\includegraphics[width=0.7\linewidth]{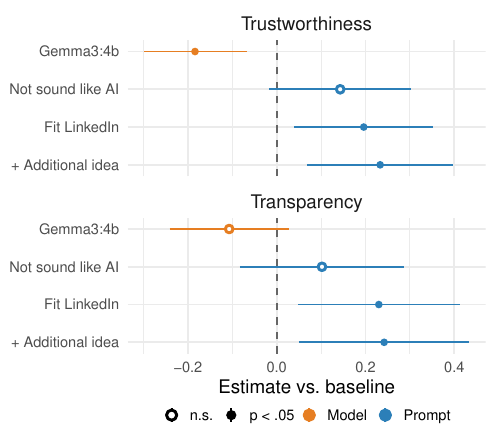}
\caption{Model and Prompt Effects. LMEM estimates (points) and 95\% CIs (bars) for AI model (reference: \texttt{Claude Sonnet 4.6}, in orange) and prompt (reference: ``Refine this post.'', in blue) on perceived trustworthiness (top) and transparency (bottom). Solid points represent statistically significant estimates ($p < 0.05$)}
\label{fig:modelprompt}
\end{figure}

To study whether trustworthiness and transparency are perceived differently across models and prompts, we run additional LMEMs as in Equation~(\ref{LMEM}) where the condition is the AI model (resp. the prompt) used to alter the post. The estimates of the fixed effects of these two LMEM are presented in Figure~\ref{fig:modelprompt}
We found that posts altered with \texttt{Gemma3:4b} were rated significantly lower in perceived trustworthiness than posts altered with \texttt{Claude Sonnet 4.6} ($b = -0.184$, $p = 0.002$, 95\% CI $[-0.299, -0.068]$). Regarding transparency, \texttt{Gemma3:4b} also scored lower but not significantly ($b = -0.107$, $p = 0.119$, CI $[-0.241, 0.027]$).
The 20 pilot participants are also part of this analysis sample; as a robustness check, we re-ran this model excluding them, with essentially unchanged estimates (Appendix, \emph{Model Comparison: Pilot-Exclusion Robustness Check}).

For prompts (reference: ``Refine this post.''), a clear pattern emerges: the more complex the prompt, the higher the ratings.
The prompts ``Fit LinkedIn'' ($b=0.196$, $p=0.015$) and ``additional idea'' ($b=0.233$, $p=0.006$) score significantly higher on both trustworthiness ($b=0.233$, $p=0.006$) and transparency ($b =0.242$, $p=0.014$). The prompt ``not sound like AI'' do not show any significant difference in both trustworthiness ($p=0.081$) and transparency ($p=0.281$) compared to the reference prompt.

\subsubsection{Each Configuration against the Human Baseline}
Against the human-written baseline (rather than each other), neither models shifts ratings significantly, with similar average absolute shift per post (trustworthiness: 0.245 vs.\ 0.257 scale points; transparency: 0.281 vs.\ 0.277; both $p>0.42$), though \texttt{Claude Sonnet 4.6} sits slightly above and \texttt{Gemma3:4b} slightly below.
By prompt, only the basic refine prompt trends (non-significantly) below the human originals on both scales; more complex prompts trend higher, though none reaches significance alone.

\subsubsection{Is There a Best Combination? Model$\times$Prompt Interaction}
The above treats model and prompt as separate, additive effects. We also tested whether they interact, on AI-altered posts only: a likelihood-ratio test found no significant model$\times$prompt interaction for either trustworthiness or transparency (both $p>.7$).
We therefore do not single out a ``best'' combination, but the eight individual model$\times$prompt cells still vary against the human baseline (full estimates in \emph{All Eight Model$\times$Prompt Cells vs.\ the Human Baseline}, Appendix): only \texttt{Claude Sonnet 4.6} with the additional-idea prompt is significant on both outcomes.

\subsection{Follow-Up Questions.}

After the two main parts, participants answered follow-up questions about on what basis they made their judgments.

\subsubsection{Judgment Factors}
Asked what they based their trustworthiness/transparency judgments on, participants mainly cited content (96\% agree/strongly agree), then writing style (79\%), whether a post sounded like AI (64\%), and formatting (61\%); topic familiarity (53\%) and links/URLs (50\%) were about half, and only 26\% relied on length (full distribution in Appendix, \emph{Follow-Up Questions}).

\subsubsection{Judgment Cues}
As an exploratory note, participants who agreed in the follow-up questions that they judged posts based on whether they sounded like AI, rated the AI-altered posts lower than the human-written ones (trustworthiness $r = -0.176$, $p = 0.011$; transparency $r = -0.153$, $p = 0.027$), while participants who disagreed rated them slightly higher.
Suspicion was not manipulated in our design, so this is a correlation and not a causal effect.

\FloatBarrier
\section{Discussion}

\subsection{Spectrum of AI in Social Media Posts}

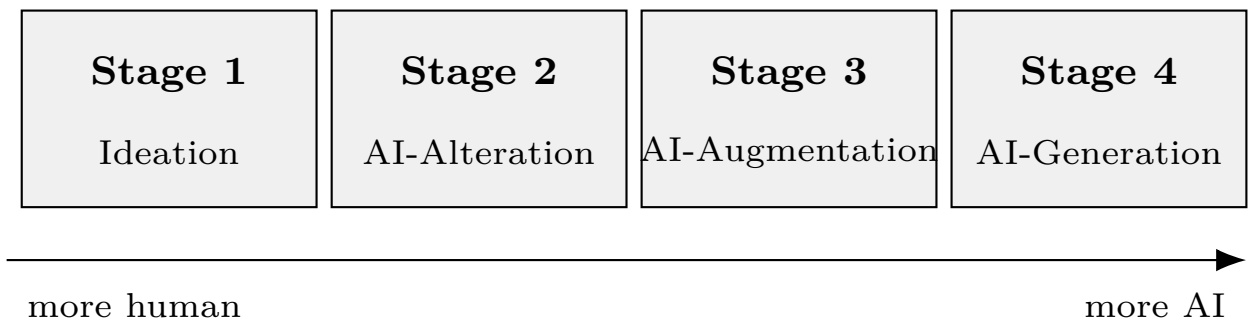
\begin{figure}[!htbp]
\centering
\begin{tikzpicture}[x=1cm, y=1cm, every node/.style={font=\fontsize{6.5}{7.5}\selectfont}]
  \def\w{2.05}
  \def\h{1.3}
  \foreach \i/\name/\desc in {0/{Stage 1}/{Ideation}, 1/{Stage 2}/{AI-Alteration}, 2/{Stage 3}/{AI-Augmentation}, 3/{Stage 4}/{AI-Generation}} {
    \pgfmathsetmacro{\x}{\i*\w}
    \draw[fill=gray!12] (\x,0) rectangle ++(\w-0.1,\h);
    \node[align=center, font=\fontsize{7.5}{8.5}\selectfont\bfseries] at (\x+0.5*\w-0.05,\h*0.68) {\name};
    \node[align=center] at (\x+0.5*\w-0.05,\h*0.28) {\desc};
  }
  \draw[-{Latex[length=2mm]}] (-0.1,-0.35) -- (3*\w+\w-0.1,-0.35);
  \node[anchor=west] at (-0.1,-0.65) {more human};
  \node[anchor=east] at (3*\w+\w-0.1,-0.65) {more AI};
\end{tikzpicture}
\caption{The spectrum of AI involvement in social media posts. This paper's prompts cover stages 2 (prompts 1--3) and 3 (prompt 4).}
\label{fig:spectrum}
\end{figure}

Related work and our own WildChat analysis show AI used for social media posts in many ways: refining language (grammar, spelling, rephrasing), changing tone or length, fitting a platform, or generating a full post from a single idea.
We organize this as a four-stage spectrum (Figure~\ref{fig:spectrum}), since AI use is a continuum rather than a human/AI binary.
In stage 1, AI is used only before writing (brainstorming, outlining), and the post itself is human-written.
Stage 2, AI-alteration, starts from a human draft and changes only its form (grammar, phrasing, tone, platform fit); our prompts 1--3 belong here.
Stage 3 adds content (an idea, question, or line) invented by the model rather than supplied or verified by the human author, while most of the post still comes from the human; our prompt 4, which also changed the posts the most, belongs here.
Stage 4, AI-generation, expands minimal human input into a full post.
The stages transition into each other, from AI ``modifying, augmenting, or generating'' a message \citep[p.~90]{hancock2020}, with alteration ending once changes become too substantial \citep[p.~302]{furey2025}; human-written posts sit outside the spectrum entirely, since no AI was involved.

The more complex the prompt (moving further into stage 2, or into stage 3), the higher the perceived trustworthiness and transparency, with the basic ``Refine this post.'' prompt scoring lowest, closest to human-written posts.
More broadly, the spectrum gives researchers and platforms a shared vocabulary for the middle ground between fully human-written and fully AI-generated posts that this paper focuses on.

\subsection{No Difference of AI in Ratings}

Neither trustworthiness nor transparency differed between human-written and AI-altered posts, with narrow confidence intervals around both null estimates, contrary to our pilot ($n=20$), which had pointed toward AI-altered posts scoring higher on both scales, and which the main study corrects.

Two patterns help explain this null result.
First, participants reported judging posts mainly by content (96\%) and least by length (26\%), while our metrics show that AI-alteration expands and reformats posts (Table~\ref{tab:metricstats}) but leaves content-level signals (TTR, sentiment direction, semantic similarity) largely intact: AI changes the form participants rely on least, while leaving intact the content they rely on most.
Second, which of the six posts a participant saw explained far more variance than condition, model, or prompt (Appendix, \emph{Exploratory Analyses: Post-Level Variance}); with only six posts, we read this only as a bound on how much condition can matter relative to other differences between posts.

Our design also differs from \citet{jakesch2019}, whose ``Replicant Effect'' (suspicion of AI authorship reducing trust) appeared only when participants believed they were rating a \emph{mixed} set of AI and human profiles (all in fact human-written), not when told all profiles were AI-generated; the effect requires readers to actively ask who wrote a text, which our unlabeled, genuinely AI-altered design gives them little reason to do.
Consistent with this, participants who self-reported judging posts by whether they ``sounded like AI'' did rate AI-altered posts lower, but since detection was only slightly better than chance, this suspicion-driven penalty rarely triggers in practice.

\subsection{The Model and Prompt Matter More}

While AI-alteration made no difference on average (H1, H2), AI-altered posts were not rated the same among each other, and this is not a contradiction.
Some model$\times$prompt combinations push ratings above the human baseline, others below (Appendix, \emph{All Eight Model$\times$Prompt Cells vs.\ the Human Baseline}), and averaging across all of them, which is exactly what H1 and H2 test, cancels these out to near zero.
The general null therefore describes AI-alteration \emph{on average}; specific choices of model and prompt still move ratings, just not consistently in one direction, meaning a given model and prompt combination can reliably raise or lower how trustworthy and transparent a post reads even though AI-alteration as a whole cannot.
\texttt{Gemma3:4b}-altered posts were rated significantly less trustworthy than \texttt{Claude Sonnet 4.6}'s. Similar patterns were observed in transparency but were not statistically significant.
Prompts show a clearer ordering: richer prompts scored higher on both scales, with only the basic ``Refine this post.'' trending below the human originals, read as a general tendency given the non-significant model$\times$prompt interactions.
The additional-idea prompt is a partial exception worth flagging on its own: it adds content invented by the model, unverified by the human author, yet it raised both trustworthiness and transparency ratings, unsurprising since readers cannot see where a post's ideas come from.
Detectability moves somewhat independently: the additional-idea prompt was detected most and the basic prompt least, while \texttt{Gemma3:4b} was harder to detect than \texttt{Claude Sonnet 4.6}, plausibly because Claude's larger, commercial output is more polished but also more recognizably ``AI-sounding.''
Model and prompt thus give authors two, not fully aligned, levers: richer prompts raise trustworthiness and transparency but also detectability, while the less-trusted model is the harder one to detect.

\subsection{Readers Cannot Detect AI-Alteration}

Detection was near chance overall, and asking the model not to sound like AI backfired: those posts were flagged as AI \emph{more} often (44.6\%) than posts from the simple refine prompt (38.7\%; Table~\ref{tab:detbycell}).
Alteration strength did not predict detection (cosine similarity to the original had no effect), but the model did, with \texttt{Claude Sonnet 4.6} flagged as AI more often than \texttt{Gemma3:4b} across every prompt, so the cue readers use is tied to a model's style, fitting that readers never see the original for comparison.
This echoes \citet{barrot2025}, on automatic detectors struggling with AI-\emph{edited} (vs.\ AI-generated) text, and \citet{dutta2026large}, whose in-the-wild audit finds such detectors produce substantial false positives even on fully AI-generated text. We show the same failure in human readers, and since our participants were explicitly asked to judge authorship (unlike a real feed, where nothing prompts the question absent suspicion), our detection rate is plausibly an upper bound.
Since readers cannot detect AI-alteration unaided, transparency about it can only come from the author or the platform, e.g., via labels or interface design \citep{gamage2025,chen2025,li2024}.

\subsection{Implications}

Prior work mostly tests the effect of disclosure: \citet{jakesch2019} found that labeling entirely human-written content as AI-assisted lowered trust (the ``replicant effect''), while \citet{li2024} found no significant effect of an AI-generated label on real AI-generated and human-written content alike; neither varies whether AI actually produced the text. Our unlabeled, genuinely AI-altered posts extend \citeauthor{li2024}'s null finding to undisclosed AI involvement itself.
Studies should keep the two questions apart and locate where on the spectrum their manipulation sits: measuring AI's effect itself calls for unlabeled, genuinely AI-altered content (as here), while measuring disclosure's effect calls for labeled content.
We also contribute reusable instruments, the adapted ABI scale and a new transparency scale, open to validation elsewhere.

For authors and platforms, using AI to refine a post carries no default penalty, but model and prompt choice pulls in different directions: richer prompts raise both perceived qualities while also raising detectability, and the less-trusted model was harder to detect.
Prompts that add new content deserve separate treatment from ones that only restyle existing text: the former can raise a post's perceived credibility without adding any actual verification, since readers cannot tell where the added material came from.
Readers should not lean on writing style alone for authorship judgments, and platforms, since readers cannot see AI involvement unaided, must decide whether and how to make it visible (grammar correction and full generation are not equivalent disclosures).

\subsection{Limitations and Future Work}

The power simulation's variance components came from only $N=20$ pilot participants, so its targets were imprecise, though the main study still reached adequate power.
Our stimuli are limited to six posts from two topic areas, two models, and four prompts with redacted authorship; future work could vary these, and author identity, to map what drives perceived trustworthiness and transparency more broadly.
Evaluating posts separately matches real feed browsing but likely understates effects a side-by-side comparison would reveal.
Finally, a rating is not behavior: future work could test whether these effects surface in liking, commenting, or sharing.
Overall, the more useful question is not whether AI was involved but how: the proposed spectrum offers a shared way to ask it.

\section{Conclusion}

We opened with a tension: AI-refined posts may look better, but readers find it harder to judge whether a post is honest or persuasive.
Our results resolve this in an unexpected direction: readers could not tell a LinkedIn post was AI-altered, and posts that raised no suspicion were rated like human-written ones, so perception is shaped by the model and the prompt, neither visible to the reader; whether AI was used at all left no detectable trace (which specific post a participant saw moved ratings by even more, though we read that only as an exploratory bound given our six stimuli).
The null H1/H2 result is itself informative: AI-alteration occupies its own place on the spectrum, distinct from AI-generation.
We contribute (1) a four-stage spectrum of AI involvement in social media posts with clear terminology, (2) evidence that AI-alteration itself leaves perceived trustworthiness and transparency unchanged while model and prompt choice do shift them, at the cost of detectability, (3) evidence that readers cannot tell an altered post from an unaltered one without cues, and instead take human authorship as the default, and (4) an adapted ABI scale, a new transparency scale, and a corpus of six human-written LinkedIn posts with their AI-altered versions, metrics, and ratings.

\section{Ethical Statement}

Participants were recruited on Prolific, compensated above its minimum recommendation, and could opt out at any time.
They were not told before the study that some posts were AI-altered, since this would have changed exactly the judgment we measured, but were fully debriefed afterward.
The study was assessed against applicable national human-subjects research regulations and confirmed to fall outside their scope (no institutional review required), no sensitive personal data was collected, and all stimuli were public LinkedIn
 posts from before 2020 with author names redacted.

Because undisclosed AI-altered posts read as trustworthy as human-written ones, the prompt that raised ratings most without adding any verification (the additional-idea prompt) could let bad actors make fabricated content look equally credible; the same prompts and model comparisons could be reused to craft posts that maximize perceived trustworthiness while evading AI-detection, e.g., for astroturfing or deceptive marketing. The prompts we test are simple restatements of patterns already common in real usage (\emph{Formative Analysis}), and our findings argue for platform-level disclosure as the actual safeguard (\emph{Implications}).

We will publicly release our stimuli corpus, alteration prompts, and analysis code upon acceptance, with post URLs stripped alongside author names, since the raw scrape (unlike the redacted stimuli shown to participants) links to real LinkedIn profiles. We did not seek individual consent from the original post authors, consistent with using only public, pre-2020 posts; a keyword screen of the corpus turned up no offensive language beyond incidental mild wording (e.g., ``hell'').

\noindent\textbf{Generative AI Use.} Large language models are experimental materials here (full names, versions, and prompts in \emph{Stimuli} and \emph{Model Screening}); separately, an AI writing assistant helped edit and restructure this manuscript's prose under full author direction, and the authors take full responsibility for all content and claims.


\FloatBarrier
\bibliographystyle{plainnat}
\bibliography{aaai2026}

\appendix
\section{Appendix}

This appendix reports additional related work moved from the main text for space, details and screenshots from the experiment, the post-level exploratory analysis that followed the pre-registered null result (\emph{Results}), and supplementary tables and figures referenced there in condensed form, for readers who want full detail.

\subsection{Additional Related Work}

\subsubsection{Generative AI, LLMs, and Writing}
Generative AI describes systems that create new content (text, images, audio) rather than only classify or predict; LLMs are GenAI models specific to language, trained to predict the next token, letting them write, summarize, translate, or rewrite text \citep{brown2020,liu2025}.
Besides commercial assistants, smaller open models such as Gemma run locally, so altering text with AI is easy and available to anyone online.
Writing assistance is a large part of this use.
\citet[p.~5]{veselovsky2023} estimate that 33--46\% of Amazon Mechanical Turk workers used LLMs for a paid text-summarization task.
In the WildChat dataset of almost one million real ChatGPT conversations, assisting with and creating text was the most common request type \citep[p.~3]{zhao2024}.
\citet{dhillon2024} found that AI sentence-level suggestions did not improve writing quality, but paragraph-level suggestions improved both quality and productivity, at a cost of lower felt ownership and satisfaction: how much an AI contributes changes the result, which motivates distinguishing degrees of AI involvement.


\subsubsection{AI Adoption and Effects on Social Media}
Beyond the adoption figures in the Introduction, the WildChat corpus itself shows users asking ChatGPT to rewrite and refine posts for different platforms \citep[p.~1]{zhao2024}, confirming AI-alteration is already common practice.
Effects are visible on platforms too: AI-generated images changed production and engagement on Pixiv \citep{wei2024,pixiv2026}; AI summaries can improve deliberation but reduce engagement \citep{heseltine2026}; and a 680-participant experiment across several AI interventions found that AI tools increase engagement and output but decrease perceived quality and authenticity \citep{moller2026}.

\subsubsection{Trustworthiness, Transparency, and AI on Social Media Posts}
Platform context shapes perception: on LinkedIn, posts are tied to verifiable identities \citep{guillory2012}, encouraging curated professional self-presentation \citep{dijck2013}, unevenly enough that even which free-text fields a user fills in varies systematically \citep{altenburger2017there}.
At the post level, message credibility and linguistic style affect follower trust and engagement \citep{lou2019,deng2020,davis2019}, and lexical models can flag misinformation by its distinct vocabulary \citep{antypas2021}; perception depends on the post's text itself, exactly what AI-alteration changes (style, length, wording, complexity, while the idea stays roughly the same).
On social media specifically, transparency has mainly been studied at the platform level (e.g., Reddit's moderation practices; \citealp{juneja2020}), not for individual posts.
Trust in an AI-disclosure label depends on its design \citep{gamage2025} and disclosed detail \citep{chen2025}; label presence itself lowers perceived authenticity of AI-generated content, with an implied-authenticity spillover raising authenticity judgments of nearby unlabeled content \citep{pawelczyk2026implied}.
\citet{jakesch2019} showed suspicion of AI authorship reduces trust even when every profile shown was actually human-written; \citet{moller2026} found AI on social media raises engagement while lowering perceived quality and authenticity; and detection tools work only to a limited extent on AI-edited, as opposed to fully AI-generated, text \citep[p.~11]{barrot2025}.

\subsection{Real-World AI-Alteration Prompting Groups}

Table~\ref{tab:promptgroup} lists the WildChat prompting groups underlying the summary in \emph{Formative Analysis}.

\begin{table}[!htbp]
\centering
\begin{tabular}{lr}
\toprule
Prompting Group & $n$ \\
\midrule
rewrite it as social media post & 10 \\
rewrite this post in a way & 10 \\
Persona & 8 \\
refine this (social media) post & 7 \\
rewrite/rephrase/paraphrase this post & 3 \\
rewrite this post to become \ldots{} (how) & 3 \\
make a social media post & 3 \\
improve this & 2 \\
shorten it for \ldots{} & 2 \\
Other (groups occurring only once) & 5 \\
\bottomrule
\end{tabular}
\caption{WildChat AI-Alteration Conversations by Prompting Group ($n = 53$).}
\label{tab:promptgroup}
\end{table}

\subsection{Model Screening: Per-Metric Detail}

Table~\ref{tab:screeningmodels} gives the per-metric breakdown behind the ``Overall'' figures in \emph{Formative Analysis}, for the 1{,}188-post screening corpus (four candidate models, three WildChat-derived prompts, 99 posts not used later in the experiment).
Each cell is the mean signed \% change for that metric and model; ``Overall'' (main text) is the mean of the \emph{absolute} values across all eight metrics shown here, not the mean of these signed values, since signed changes in opposite directions would otherwise cancel out.

\begin{table}[!htbp]
\centering
\begin{tabular}{lrrrr}
\toprule
Metric & Qwen2.5:7b & Gemma3:4b & GPT-5.2 & Claude 4.6 \\
\midrule
Word count & $+12.8$ & $+78.0$ & $+16.2$ & $+122.6$ \\
Special characters. & $+25.8$ & $+88.2$ & $+27.9$ & $+111.8$ \\
Punctuation & $+29.8$ & $+69.4$ & $+6.6$ & $+113.3$ \\
Paragraphs & $+5.6$ & $-8.3$ & $+58.6$ & $+119.1$ \\
Line breaks & $-1.9$ & $-18.8$ & $+33.1$ & $+76.2$ \\
TTR & $+8.6$ & $+5.2$ & $+3.1$ & $-3.2$ \\
Uppercase ratio & $+17.1$ & $-13.2$ & $+3.8$ & $-4.0$ \\
Sentiment score & $+7.1$ & $+17.0$ & $+2.8$ & $+5.9$ \\
Cosine similarity & $0.878$ & $0.800$ & $0.890$ & $0.816$ \\
\bottomrule
\end{tabular}
\caption{Screening corpus ($n=1{,}188$): mean signed \% change per metric, by model. Note that, cosine similarity is not a change score.}
\label{tab:screeningmodels}
\end{table}

\subsection{Participant Instructions}

\begin{figure}[!htbp]
\centering
\includegraphics[width=0.7\linewidth]{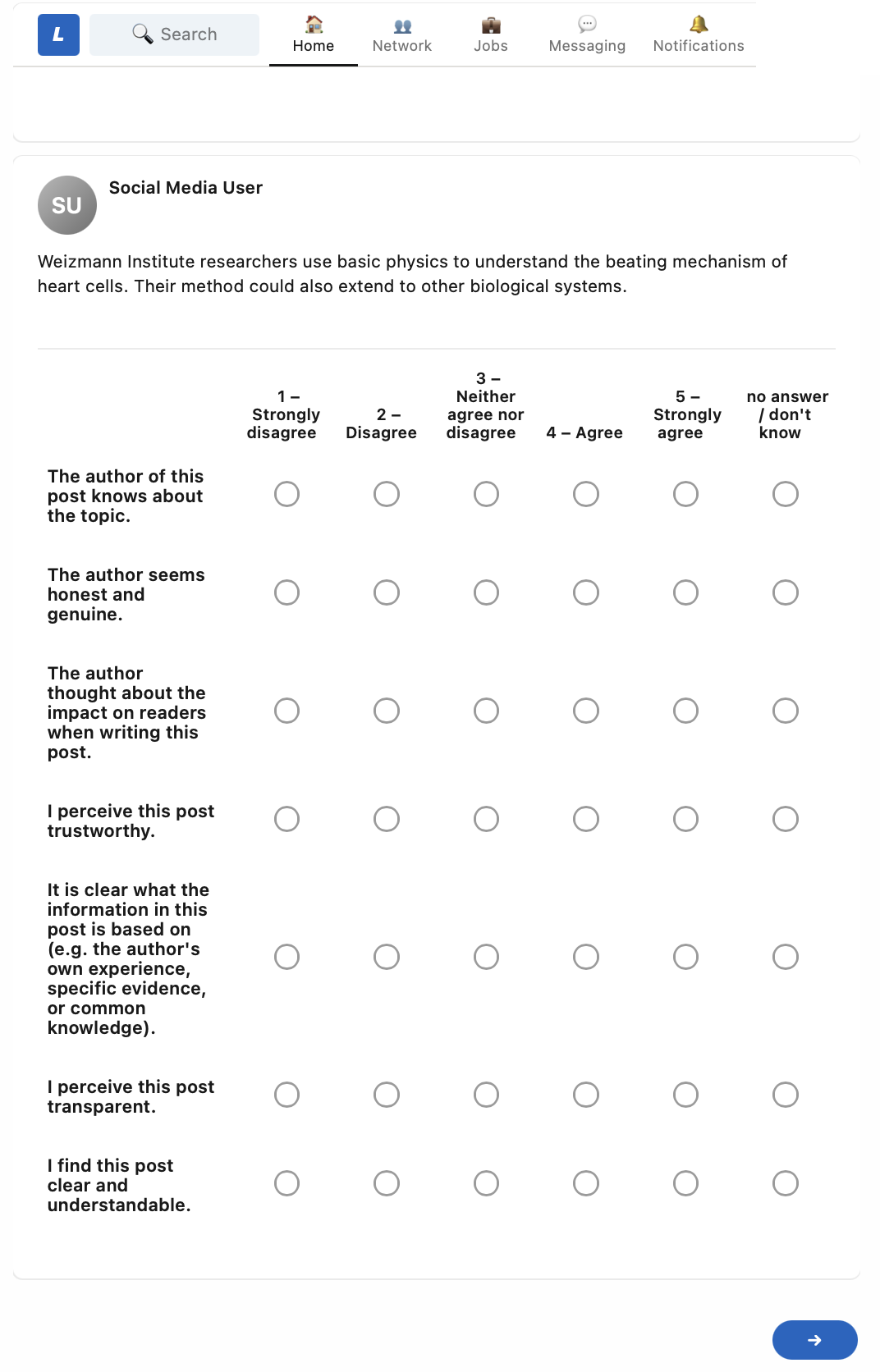}
\caption{Example rating task (Part 1), showing the trustworthiness and transparency items for one post.}
\label{fig:ratingtask}
\end{figure}

\begin{figure}[!htbp]
\centering
\includegraphics[width=0.7\linewidth]{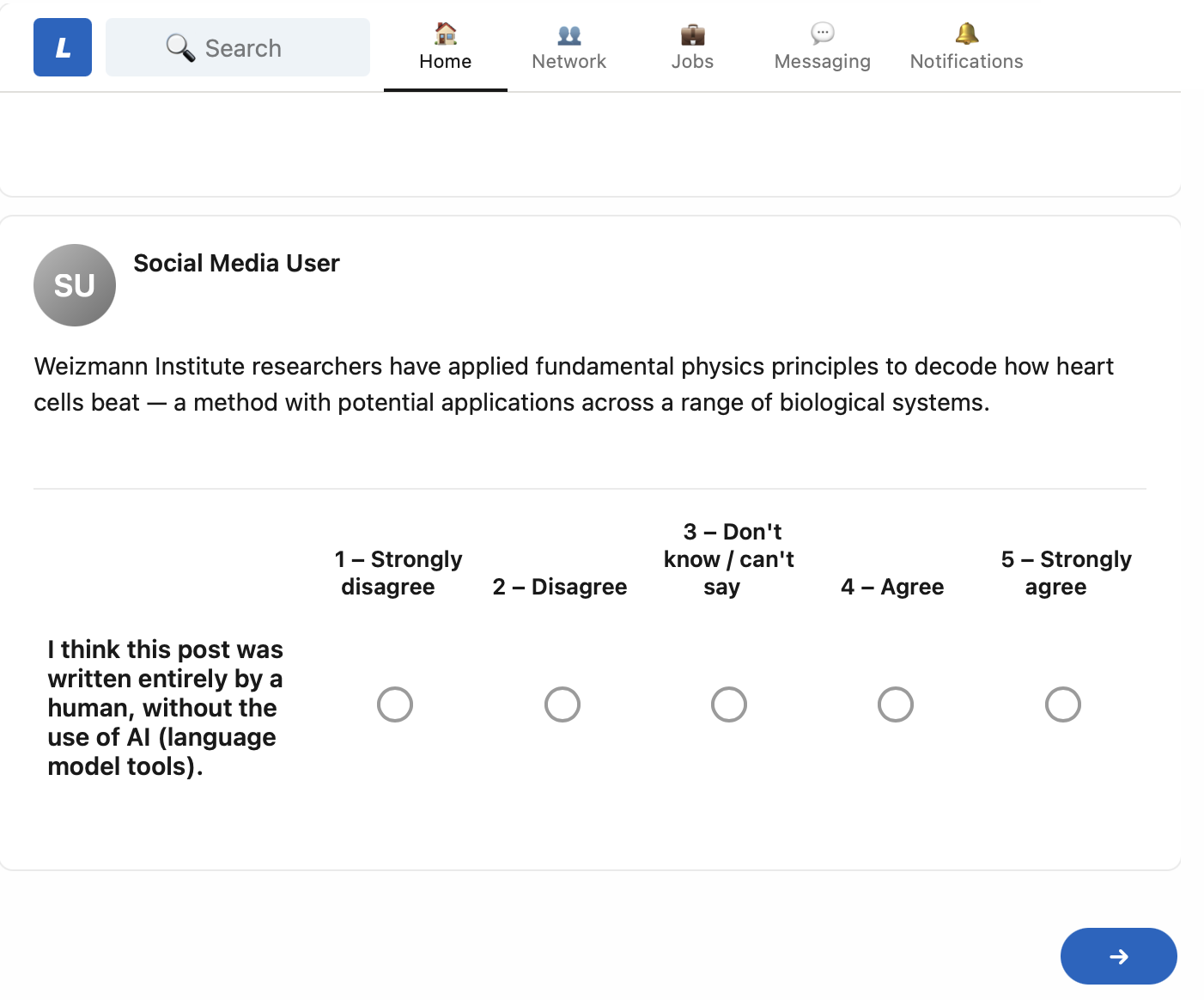}
\caption{Example AI-detection task (Part 2), rating the same post shown in Figure~\ref{fig:ratingtask}.}
\label{fig:aidetectiontask}
\end{figure}

\begin{figure}[!htbp]
\centering
\includegraphics[width=0.7\linewidth]{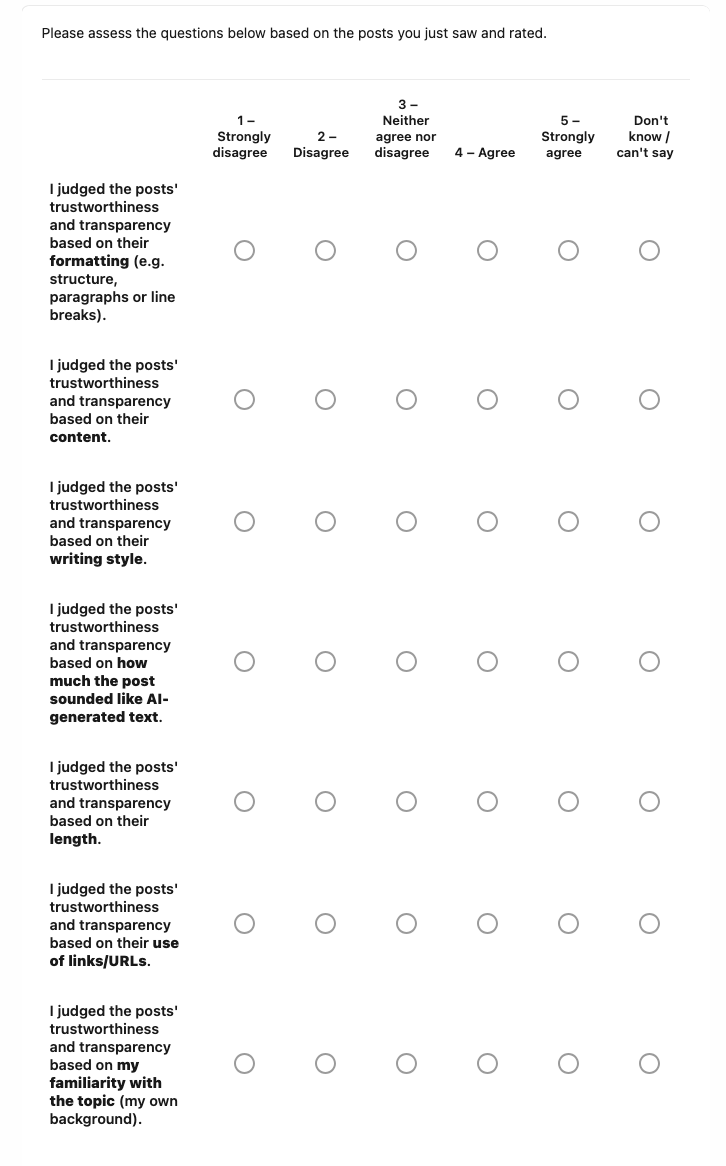}
\caption{Post-survey questions, section (i).}
\label{fig:post_survey_p1}
\end{figure}

\begin{figure}[!htbp]
\centering
\includegraphics[width=0.7\linewidth]{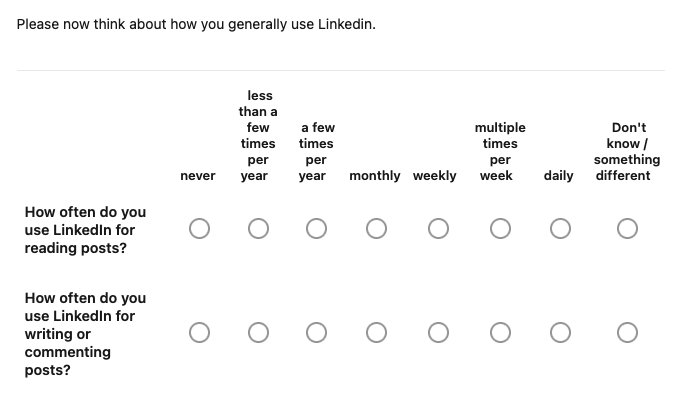}
\caption{Post-survey questions, section (ii).}
\label{fig:post_survey_p2}
\end{figure}

\begin{figure}[!htbp]
\centering
\includegraphics[width=0.7\linewidth]{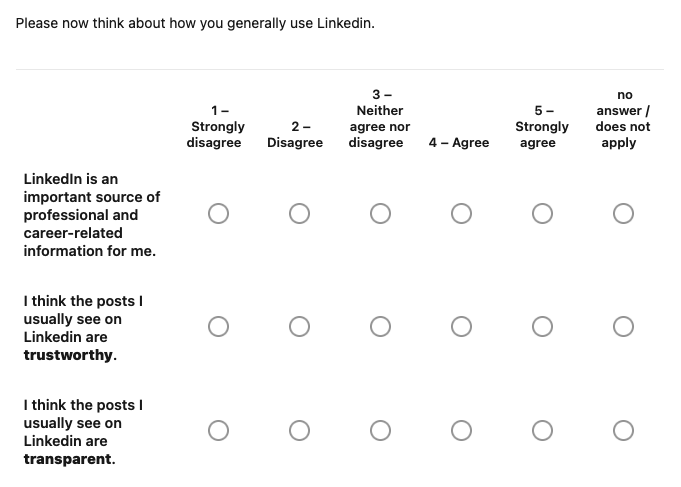}
\caption{Post-survey questions, section (iii).}
\label{fig:post_survey_p3}
\end{figure}

At onboarding, participants read: ``Welcome to the survey about trustworthiness and transparency of social media posts. You will see a series of social media posts, similar to what you'd find on LinkedIn. Your task is to assess each post. Parts 1 and 2 have 4 posts to rate, followed by Part 3 with a few follow-up questions. This study is part of an academic research project; for any questions, please message the researcher directly on Prolific.'' Figure~\ref{fig:ratingtask} shows an example rating task from Part 1, and Figure~\ref{fig:aidetectiontask} an example AI-detection task from Part 2. Figures~\ref{fig:post_survey_p1},~\ref{fig:post_survey_p2} and~\ref{fig:post_survey_p3} show the follow-up questions as they appear to the participants.

After completing all parts, participants read the following debriefing: ``Thank you for participating. Your response has been saved. [...] Your responses contribute to a research project investigating whether AI-altered social media posts are perceived as trustworthy and transparent. In Part 1, you saw an example post and 1 post written by a human and 3 posts altered using artificial intelligence (AI), randomly assigned. You rated their trustworthiness and transparency to examine whether AI-altered posts are perceived differently from human-written ones. In Part 2, you assessed the same posts to indicate whether you thought each was written by a human or altered by AI. This allows us to examine whether perceived AI authorship influences how posts are rated. Part 3 included general questions about your social media use, to explore whether factors such as usage frequency relate to how trustworthy or transparent posts are perceived.''

\subsection{Post-Level Metrics vs.\ Ratings Correlations}

Besides the pre-registered hypotheses, we tested whether the text metrics of the altered posts explain the ratings.
Using the same model structure as in \emph{Analysis Plan and Statistical Model}, we related the cosine similarity to the human original, the direction of the change in lexical diversity (TTR), the post length, the average sentence length, and the direction of the sentiment change to perceived trustworthiness and transparency.
None of them explained the ratings.
How strongly a post was altered, and in which linguistic direction, was unrelated to how it was rated.
The only estimate that reached significance was the sentiment direction on transparency ($b = 0.361$, $p = 0.023$), which rests on four posts only and points in the opposite direction of what we expected.
The same holds for detection: the cosine similarity had no effect on whether a post was flagged as AI, also not when the model was held constant.
These analyses are exploratory and descriptive only.

To make this concrete, we correlate (Pearson, $n=48$ AI-altered stimuli) each post-level metric with mean trustworthiness/transparency, given in full in Table~\ref{tab:metriccorr}: cosine similarity ($r=.157$/$.107$), TTR \% change ($r=-.141$/$-.145$), word count \% change ($r=-.069$/$-.023$), words per sentence ($r=.079$/$.163$), and sentiment score \% change ($r=.010$/$.098$).
The intuition going in was that a more heavily altered post (lower cosine similarity, larger TTR or length change) might read as less authentic and score lower; instead, none of the correlations approach significance (all $p>.26$), echoing the LMEM result that alteration strength does not predict perceived trustworthiness or transparency.

\begin{table}[!htbp]
\centering
\small
\begin{tabular}{lrr}
\toprule
Metric ($n=48$) & $r$ (Trust) & $r$ (Transp.) \\
\midrule
Cosine similarity & 0.157 & 0.107 \\
TTR (\%) change & $-0.141$ & $-0.145$ \\
Word count (\%) change & $-0.069$ & $-0.023$ \\
Words per sentence & 0.079 & 0.163 \\
Sentiment score (\%) change & 0.010 & 0.098 \\
\bottomrule
\end{tabular}
\caption{Pearson correlations between post-level alteration metrics and mean trustworthiness/transparency ratings across the 48 AI-altered stimuli. None reach $p<.05$ (all $p>.26$).}
\label{tab:metriccorr}
\end{table}

\subsection{Exploratory Analyses: Post-Level Variance}

\begin{figure}[!htbp]
\centering
\includegraphics[width=0.7\linewidth]{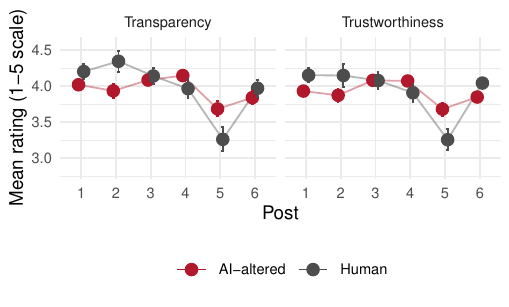}
\caption{Mean trustworthiness and transparency ratings per post (1--6), split by human-written vs.\ AI-altered. Error bars are $\pm 1$ SE.}
\label{fig:perpostratings}
\end{figure}

We checked how much of the difference in ratings comes from which of the six posts a participant saw (regardless of human/AI condition), versus condition, model, or prompt.
Which post explains far more (4.95\% of trustworthiness variance, 4.88\% transparency) than human-vs-AI (0.08\%/0.07\%), AI model (0.96\%/0.26\%), or prompt (0.67\%/0.81\%).
Post 5 (short, two sentences, no personal statements) scores lowest on trustworthiness and transparency of all six (Figure~\ref{fig:perpostratings}), and per-post detection accuracy ranged from 37.5\% to 67.7\% correct.
With only six posts drawn from two topic areas, we cannot attribute this to any generalizable property of a post (topic, length, tone, or otherwise): a larger and more topically diverse pool of stimuli would be needed to test whether post-level differences trace to something systematic rather than being idiosyncratic to these six items. We therefore read this result only as showing that which post a participant happens to see moves ratings more than our experimental manipulations do, not as a claim about what kind of post is more or less trusted.

\begin{figure}[!htbp]
\centering
\includegraphics[width=0.7\linewidth]{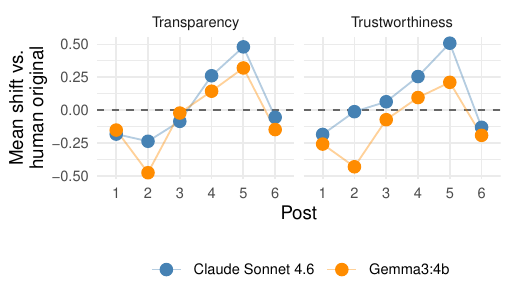}
\caption{Main Study: Mean Shift From the Human Original per Post and Model, Averaged Over the Four Prompt Variants.}
\label{fig:shifts}
\end{figure}

\textbf{Shifts per Post.}
Figure~\ref{fig:shifts} shows how much the perception of transparency and trustworthiness in AI-altered posts differs from the ones of the human posts, grouped by AI model.
Two patterns stand out: the direction of the shift depends on the post, not the model (both models score lower on posts 1, 2, 6 and higher on posts 4, 5, by up to 0.5 scale points, larger than the overall averages, so the H1/H2 nulls hold on average but not post-by-post), and \texttt{Claude Sonnet 4.6} scores consistently, though slightly, higher than \texttt{Gemma3:4b} across almost all posts (we did not model this at the post level given too few ratings per cell).

\subsection{Model Comparison: Pilot-Exclusion Robustness Check}

Table~\ref{tab:robustness} reports the AI-model LMEM (trustworthiness/transparency $\sim$ ai\_model + (1|participant) + (1|post)) fit with and without the 20 pilot participants, referenced in \emph{AI Models and Prompts}.

\begin{table}[!htbp]
\centering
\small
\begin{tabular}{lrrr}
\toprule
Sample & Outcome & Estimate & $p$ \\
\midrule
With pilot ($n=214$) & Trust & $-0.184$ & .0019 \\
With pilot ($n=214$) & Transp. & $-0.107$ & .1186 \\
No pilot ($n=194$) & Trust & $-0.200$ & .0014 \\
No pilot ($n=194$) & Transp. & $-0.130$ & .0697 \\
\bottomrule
\end{tabular}
\caption{\texttt{Gemma3:4b} vs.\ \texttt{Claude Sonnet 4.6} estimate, with and without the 20 pilot participants.}
\label{tab:robustness}
\end{table}

\subsection{All Eight Model$\times$Prompt Cells vs.\ the Human Baseline}

Table~\ref{tab:eightcells} gives the full per-cell estimates underlying \emph{Is There a Best Combination?}, from one LMEM per outcome ($\mathit{outcome} \sim \mathit{cell} + (1|\mathit{participant}) + (1|\mathit{post})$) with the human-written posts as the reference cell.

\begin{table}[!htbp]
\centering
\small
\begin{tabular}{lrrrr}
\toprule
& \multicolumn{2}{c}{Trust} & \multicolumn{2}{c}{Transp.} \\
Cell (vs.\ human) & Est. & $p$ & Est. & $p$ \\
\midrule
Claude + refine & $-0.093$ & .315 & $-0.096$ & .365 \\
Claude + not AI & $+0.110$ & .255 & $+0.033$ & .765 \\
Claude + fit LI & $+0.132$ & .160 & $+0.100$ & .356 \\
Claude + +idea & $+0.249$ & \textbf{.012} & $+0.228$ & \textbf{.046} \\
Gemma + refine & $-0.174$ & .067 & $-0.178$ & .104 \\
Gemma + not AI & $-0.041$ & .651 & $-0.056$ & .593 \\
Gemma + fit LI & $+0.024$ & .799 & $+0.100$ & .352 \\
Gemma + +idea & $-0.071$ & .493 & $-0.006$ & .956 \\
\bottomrule
\end{tabular}
\caption{All eight model$\times$prompt cells vs.\ the human baseline. LI = LinkedIn.}
\label{tab:eightcells}
\end{table}

\subsection{Follow-Up Questions}

\begin{figure}[!htbp]
\centering
\includegraphics[width=0.75\textwidth]{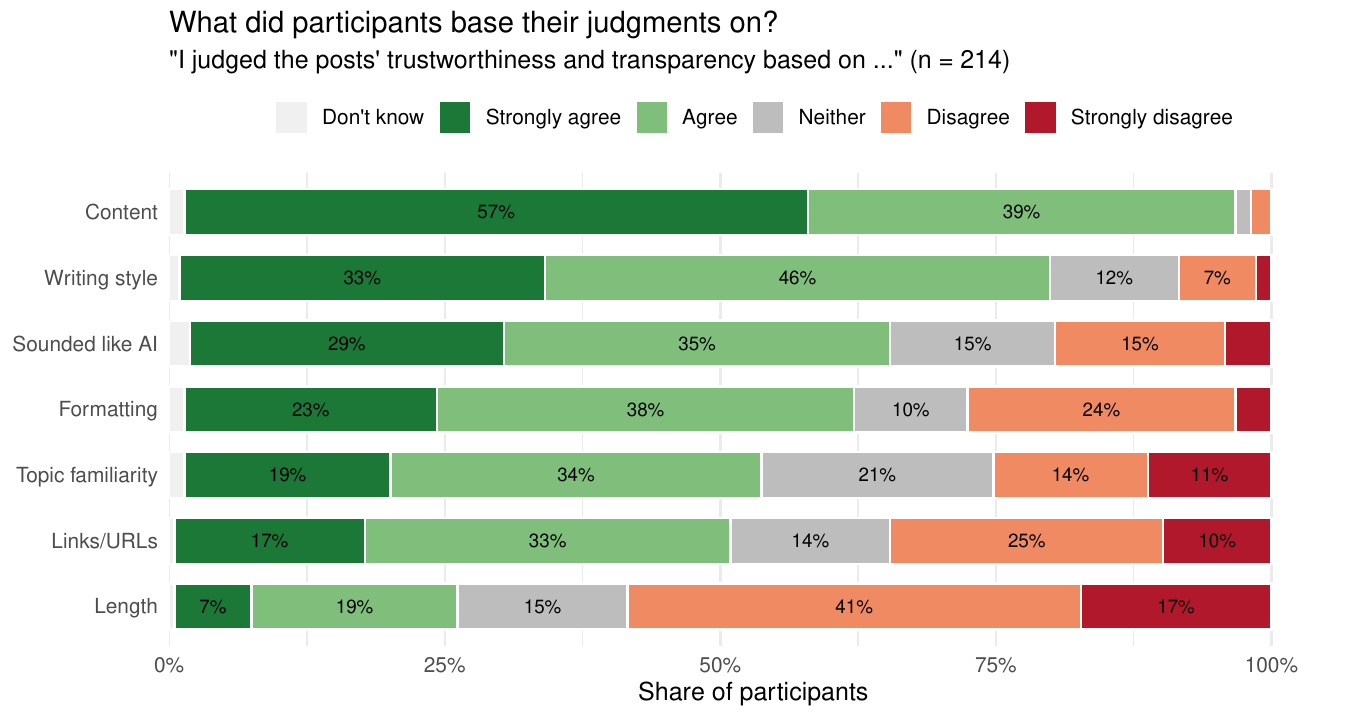}
\caption{Main Study: Self-Reported Factors for Trustworthiness and Transparency Judgments.}
\label{fig:judgment}
\end{figure}

Figure~\ref{fig:judgment} shows the full distribution behind the \emph{Judgment Factors} summary in \emph{Follow-Up Questions}.

\subsection{Demographics of participants.}

The combined sample ($n=214$) is young (mean age 32.1, 72\% aged 19--35) and predominantly male (61\%), from 33 countries (mostly Africa 43\%, Europe 28\%, Asia 20\%; South Africa, Egypt, and India together about half); 37\% are English native speakers and 45\% employed full-time.
All use LinkedIn (a recruitment criterion) and most also use YouTube (94\%), Facebook (81\%), and Instagram (77\%); we aimed for representative LinkedIn users, not a global-population sample.

\subsection{Per-Post and Per-Metric Alteration Detail (Main Experiment)}

Table~\ref{tab:perpost} gives the per-post detail behind the \emph{Manipulation Check} summary, averaged over each post's eight AI-altered versions; Table~\ref{tab:experimentmetrics} gives the per-metric detail by model and by prompt referenced in the same section, for the 48 main-experiment AI-altered posts.

\begin{table}[!htbp]
\centering
\begin{tabular}{lrrr}
\toprule
Post & Overall & Sentiment & Cos Sim \\
\midrule
Post 1 & $+14.19$\% & $+10$\% & 0.925 \\
Post 2 & $+13.79$\% & $+3$\% & 0.855 \\
Post 3 & $+9.71$\% & $-13$\% & 0.819 \\
Post 4 & $+29.60$\% & $+68$\% & 0.893 \\
Post 5 & $+67.62$\% & $+14$\% & 0.839 \\
Post 6 & $+25.32$\% & $+11$\% & 0.833 \\
\bottomrule
\end{tabular}
\caption{Changes per post of the main experiment, averaged over the eight AI-altered versions of each post. ``Overall'' is the mean absolute \% change across the eight metrics in Table~\ref{tab:experimentmetrics}.}
\label{tab:perpost}
\end{table}

\begin{table}[!htbp]
\centering
\small
\begin{tabular}{lrrrrrr}
\toprule
& \multicolumn{2}{c}{Model} & \multicolumn{4}{c}{Prompt}\\
Metric & Claude & Gemma & Refine & Not AI & Fit LI & +idea \\
\midrule
Word count & $+64.3$ & $+19.1$ & $+4.0$ & $+20.6$ & $+28.8$ & $+113.3$ \\
Special chars. & $+113.7$ & $+80.8$ & $+20.8$ & $+68.6$ & $+84.5$ & $+215.1$ \\
Punctuation & $+90.4$ & $+43.2$ & $+4.4$ & $+46.3$ & $+63.2$ & $+153.4$ \\
Paragraphs & $+174.3$ & $+9.4$ & $+46.5$ & $+46.8$ & $+105.8$ & $+168.3$ \\
Line breaks & $+94.6$ & $+33.3$ & $+27.9$ & $+21.3$ & $+78.7$ & $+128.0$ \\
TTR & $-1.5$ & $+2.9$ & $+1.4$ & $+3.4$ & $+3.5$ & $-5.5$ \\
Uppercase ratio & $+6.2$ & $+11.7$ & $+3.5$ & $+4.2$ & $+16.5$ & $+11.6$ \\
Sentiment score & $+2.2$ & $+29.3$ & $+9.8$ & $+15.9$ & $+15.7$ & $+21.6$ \\
\bottomrule
\end{tabular}
\caption{Main experiment ($n=48$): mean signed \% change per metric, by model and by prompt. Not AI = ``not sound like AI''; Fit LI = ``fit LinkedIn''; +idea = additional-idea prompt.}
\label{tab:experimentmetrics}
\end{table}
\FloatBarrier
\end{document}